\documentstyle[12pt,a41,epsfig]{article}

\newcommand{\DD}{\Delta}

\newcommand\MeV{\,\mbox{MeV}}
\newcommand\GeV{\,\mbox{GeV}}

\newcommand\Fvec{\,\mbox{\boldmath $F$}}
\newcommand\Gvec{\,\mbox{\boldmath $G$}}
\newcommand\Mvec{\,\mbox{\boldmath $M$}}
\newcommand\PV{\,\mbox{\boldmath $P$}}
\newcommand\MV{\,\mbox{\boldmath $M$}}

\newcommand\UV{\,\mbox{\boldmath $U$}}

\begin{document}
\setlength{\baselineskip}{0.52cm}
\sloppy
\thispagestyle{empty}
\begin{flushleft}
DESY 96--050 \\
WUE--ITP--96--008 \\
May 1996\\
\end{flushleft}

\setcounter{page}{0}

\mbox{}
\vspace*{\fill}
\begin{center}
{\Large\bf The Singlet Contribution to the Structure Function} \\
\vspace{3mm}
{\Large\bf  $g_1(x,Q^2)$ at Small $x$}\\

\vspace{5em}
\large
J. Bl\"umlein$^a$ and A. Vogt$^{b,c}$
\\
\vspace{5em}
\normalsize
{\it   $^a$DESY--Zeuthen}\\
{\it   Platanenallee 6, D--15735 Zeuthen, Germany}\\
\vspace{5mm}
{\it   $^b$Deutsches Elektronen-Synchrotron DESY}\\
{\it   Notkestra\ss{}e 85, D--22603 Hamburg, Germany}\\
\vspace{5mm}
{\it   $^c$Institut f\"ur Theoretische Physik,
Universit\"at W\"urzburg}\\
{\it Am Hubland,
D--97074 W\"urzburg, Germany~\footnote{Address after May 1st 1996.}
}\\
\end{center}
\vspace*{\fill}
\begin{abstract}
\noindent
The resummation of $O(\alpha_s^{l+1} \ln^{2l} x)$ terms in the
evolution equation of the singlet part of $g_1(x,Q^2)$ is carried out. 
The corresponding singlet evolution kernels are calculated explicitely. 
The leading small-$x$ contribution to the three-loop splitting function 
matrix is determined in the $\overline{\rm MS}$ scheme. Relations are 
derived for the case of ${\cal N} = 1$ supersymmetric Yang--Mills field 
theory.
Numerical results are presented for the polarized singlet and gluon 
densities, and the structure functions $g_1^{\, p}(x,Q^2)$ and 
$g_1^{\, n}(x,Q^2)$. They are compared for different assumptions on the 
non--perturbative input distributions, and the stability of the results
against presently unknown subleading contributions is investigated.
\end{abstract}
\vspace*{\fill}
\newpage
%
%
%%%%%%%%%%%%%%%%%%%%%%%%%%%%%%%%%%%%%%%%%%%%%%%%%%%%%%%%%%%%%%%%%%%%%%%%
\section{Introduction}
\label{sect1}
%%%%%%%%%%%%%%%%%%%%%%%%%%%%%%%%%%%%%%%%%%%%%%%%%%%%%%%%%%%%%%%%%%%%%%%%
%
%
The small--$x$ behaviour of polarized structure functions is a largely
unexplored subject. The current measurements cover at most the range 
$ x > 5 \cdot 10^{-3}$~\cite{VOSS}. At even lower $x$-values both size 
and sign of the structure functions $g_1^{\, p}(x,Q^2)$ and $g_1^{\, n}
(x,Q^2)$ are yet unknown. The initial distributions at a starting scale 
$Q_0^2$ of the QCD evolution can not be determined within perturbative 
QCD. Irrespectively of the specific
behaviour of these non--perturbative 
quantities one may consider, however, the QCD evolution of the structure
functions and parton densities. 
Besides the well--known leading order anomalous dimensions~\cite{LO}, 
recently also the next--to--leading order ones have been calculated 
\cite{NLO}. Furthermore the all--order resummation of leading singular 
terms in higher--order anomalous dimensions may be important in the
small-$x$ range. These terms behave as $ \alpha_s^{l+1} \ln^{2l} x $, 
corresponding to $ N(\alpha_s/N^2)^{l+1} $ for $ N \rightarrow 0 $
in the Mellin moment plane. Their numerical effect on the evolution of
non--singlet~\cite{KL} structure functions has  been analyzed in
refs.~\cite{BV1,BV2}. Contrary to an earlier expectation in ref.~\cite
{BER1} it turns out to be very
small and is found to depend on the way in 
which fermion number conservation is imposed, i.e.\ also
on less singular
 $\ln^k x$ contributions. Recently also an equation for the
leading small-$x$ resummed singlet evolution has been obtained 
\cite{BER2}.

In the present paper we study this singlet resummation in the framework 
of the renormalization group equation. After recalling the general 
evolution equation and setting up our notations in Section~2, we 
calculate the anomalous dimension matrix as a series in $\alpha_s$ 
explicitely in Section~3. A series of properties of this matrix is 
discussed. The numerical effect of the resummation beyond the known 
\cite{NLO} next--to--leading order effects on the small-$x$ behaviour 
of the polarized singlet quark and gluon densities, $\Delta \Sigma
(x,Q^2)$ and $\Delta g(x,Q^2)$, and on the structure functions 
$g_1^{\, p} (x,Q^2)$ and $g_1^{\, n}(x,Q^2)$ is then studied in 
Section~4, using input distributions as determined in recent analyses
\cite{GRSV}. We also investigate the stability of the results against 
yet unknown effects of terms less singular as $x \rightarrow 0$. 
Section~5 contains our conclusions.
%
%
%%%%%%%%%%%%%%%%%%%%%%%%%%%%%%%%%%%%%%%%%%%%%%%%%%%%%%%%%%%%%%%%%%%%%%%%
\section{The Evolution Equation}
\label{sect2}
%%%%%%%%%%%%%%%%%%%%%%%%%%%%%%%%%%%%%%%%%%%%%%%%%%%%%%%%%%%%%%%%%%%%%%%%
%
%
The evolution equation for the polarized singlet quark and gluon 
densities $(\Delta \Sigma, \Delta g)$ is given by
%-----------------------------------------------------------------------
\begin{equation}
\label{eq21}
 \frac{\partial}{\partial \ln Q^2}
 \left ( \begin{array}{c} \Delta \Sigma (x, Q^2) \\
                          \Delta g(x,Q^2) \end{array} \right )
 = \PV (x, \alpha_s) \otimes  \left (
 \begin{array}{c} \Delta \Sigma (x, Q^2) \\
                         \Delta g(x,Q^2)  \end{array} \right ).
\end{equation}
%-----------------------------------------------------------------------
Here $\otimes$ stands for the Mellin convolution, and the matrix of the 
polarized singlet splitting functions $\PV (x,\alpha_s) $ is specified 
below. In the following, we will simplify the notation by using the 
abbreviation $a_s \equiv \alpha_s(Q^2) /4 \pi $ for the running QCD 
coupling. The scale dependence of $a_s$ is obtained from
%-----------------------------------------------------------------------
\begin{equation}
 \label{eq22}
 \frac{da_s}{d\ln Q^2} = - \sum_{k=0}^{\infty} a^{k+2}_s \beta_k \: ,
\end{equation}
%-----------------------------------------------------------------------
where only $ \beta_0 = (11/3)\, C_A - (4/3)\, T_F N_f $ and
$ \beta_1 = (34/3)\, C_A^2 - (20/3)\, C_A T_F N_f -  4\, C_F T_F N_f$ 
enter up to next--to--leading order (NLO). Here $ C_A = N_c \equiv 3 $, 
$ C_F = (N_c^2-1)/(2 N_c) \equiv 4/3 $, $ T_F = 1/2 $, and $N_f$
denotes the number of flavours.
The matrix $\PV (x,\alpha_s)$ can be represented by the series
%-----------------------------------------------------------------------
\begin{equation}
 \label{eqPS}
 \PV (x,a_s) = \sum_{l=0}^{\infty} a_s^{l+1} \PV^{(l)}(x) \: .
\end{equation}
%-----------------------------------------------------------------------
Unlike the LO and NLO~($\overline{\rm MS}$) transition matrices 
$\PV^{(0)}(x)$ and $\PV^{(1)}(x)$ \cite{LO,NLO}, the splitting functions
$\PV^{(l>2)}(x)$ are not completely known so far. For the solution of 
eq.~(\ref{eq21}) beyond NLO in Section 4, we will use the asymptotic 
small-$x$ form of the latter matrices as derived in Section~3.

From that solution the structure function $g_1(x,Q^2)$ is finally
obtained by the convolution
%-----------------------------------------------------------------------
\begin{equation}
 \label{eq23}
 g_{1}^{\, p,n}(x,Q^2) = \frac{1}{9}  \left\{ c_{1,q}(x,Q^2) \otimes
 \Delta \Sigma(x,Q^2) + c_{1,g}(x,Q^2) \otimes \Delta g(x,Q^2) \right\}
 + g_{1,{\rm NS}}^{\, p,n}(x,Q^2)
\end{equation}
%-----------------------------------------------------------------------
with
%-----------------------------------------------------------------------
\begin{equation}
\label{eq23A}
 \Delta \Sigma = \DD u + \DD \overline{u} + \DD d
  + \DD \overline{d} + \DD s + \DD \overline{s} \: ,
\end{equation}
%-----------------------------------------------------------------------
and $ \DD u, \DD d, \DD s$ denoting the polarized up--, down--, and 
strange--quark distributions. We consider only the contribution
of the three light flavours. The non--singlet part of $g_1(x,Q^2)$ has 
been dealt with in refs.~\cite{BV1,BV2} already, to which we refer for 
further details.
The coefficient functions $c_{1}(x,Q^2)$ can be expanded in the strong
coupling as
%-----------------------------------------------------------------------
\begin{equation}
\label{eq24}
 c_{1,i}(x,Q^2) = \delta(1-x)\delta_{iq} + \sum_{l=1}^{\infty}
 a^l_s  c_{1,i}^{(l)}(x) \: .
\end{equation}
%-----------------------------------------------------------------------
Here it is important to note that in the $\overline{\rm MS}$ scheme
the known coefficient functions $c_{1}^{(l)}(x)$ for both $l = 1$ 
and $l = 2$ (cf.\ refs.~\cite{CO1,CO2}) behave only like
%-----------------------------------------------------------------------
\begin{equation}
\label{eq25}
 c_1^{(1)} \propto  \alpha_s \ln \left (\frac{1}{x} \right) \: , \:\:\:
 c_1^{(2)} \propto  \alpha_s^2 \ln^3 \left (\frac{1}{x} \right) 
\end{equation}
%-----------------------------------------------------------------------
at small $x$.
Therefore a prediction can be made on the small-$x$ behaviour of the
three--loop transition matrix in the $\overline{\rm MS}$--scheme, 
$\PV_{x \rightarrow 0}^{(2)}$, 
see eq.~(\ref{eqP2}) below.
Note that in eq.~(\ref{eq21}) the resummation under consideration is of 
leading order, i.e.\ the respective terms emerge only together with 
$\beta_0$ after rewriting the evolution equation in terms of $a_s$. 
Details of the solution will be presented elsewhere \cite{BRV}.
%
%
%%%%%%%%%%%%%%%%%%%%%%%%%%%%%%%%%%%%%%%%%%%%%%%%%%%%%%%%%%%%%%%%%%%%%%%%
\section{Resummation of dominant terms for $x \rightarrow 0$}
\label{sect3}
%%%%%%%%%%%%%%%%%%%%%%%%%%%%%%%%%%%%%%%%%%%%%%%%%%%%%%%%%%%%%%%%%%%%%%%%
%
%
The resummed transition matrix for the leading singular terms as $x 
\rightarrow 0$, $\PV (x, a_s)_{x \rightarrow 0}$, can be obtained from 
the solution of eq.~(\ref{eq31}) via inverse Mellin transformation:
%-----------------------------------------------------------------------
\begin{equation}
\label{eqSER}
 \PV(x, a_s)_{x \rightarrow 0} \equiv \sum_{l=0}^{\infty}
 \PV_{x \rightarrow 0}^{(l)} \, a_s^{l+1} \ln^{2l} x = 
 \frac{1}{8 \pi^2} {\cal M}^{-1} \left[  \Fvec_0(N, a_s)\right] (x) \: .
\end{equation}
%-----------------------------------------------------------------------
The matrix--valued function $\Fvec_0(N,a_s)$ is subject to the relation
%-----------------------------------------------------------------------
\begin{equation}
\label{eq31}
\Fvec_0(N,a_s) = 16 \pi^2 \frac{a_s}{N} \MV_0 - \frac{8 a_s}{N^2}
\Fvec_8(N,a_s) \Gvec_0
+ \frac{1}{8 \pi^2} \frac{1}{N} \Fvec_0^2(N,a_s) 
\end{equation}
%-----------------------------------------------------------------------
derived in ref.~\cite{BER2}, where $\Fvec_8(N,a_s)$ is the solution of
%-----------------------------------------------------------------------
\begin{equation}
\label{eq32}
\Fvec_8(N,a_s) = 16 \pi^2 \frac{a_s}{N} \MV_8 + \frac{2 a_s}{N} C_G
\frac{d}{d N}
\Fvec_8(N,a_s) + \frac{1}{8 \pi^2} \frac{1}{N} \Fvec_8^2(N,a_s) \: .
\end{equation}
%-----------------------------------------------------------------------
The basic colour factor matrices are given by 
%-----------------------------------------------------------------------
\begin{equation}
\MV_0 = \left ( \begin{array}{cc}   C_F & -2 T_F N_f \\
                                    2 C_F &  4 C_A     \end{array}
 \right ),~~~~~\Gvec_0 = \left ( \begin{array}{cc}
 C_F & 0 \\
 0 &  C_A     \end{array} \right ),~~~~~\MV_8
 = \left ( \begin{array}{cc} C_F - C_A/2 &  -T_F N_f \\
                             C_A &  2 C_A    \end{array}
 \right).
\end{equation}
%-----------------------------------------------------------------------
We determine the matrix $\PV(x, a_s)_{x \rightarrow 0}$ in terms of
a series in $a_s$ since this representation is needed for the solution 
of the singlet evolution equation~(\ref{eq21}), see Section~4. 

The lowest order entries read
%-----------------------------------------------------------------------
\begin{equation}
 \PV^{(0)}_{x\rightarrow 0} = 2 \left(
 \begin{array}{cc}    {\it C_F} & -2\,{\it T_F N_f}\\
                   2\,{\it C_F} &  4\,{\it C_A} \end{array} \right) ,
\label{eqP0}
\end{equation}
%-----------------------------------------------------------------------
\begin{equation}
 \PV^{(1)}_{x\rightarrow 0} = 2 \left( 
 \begin{array}{cc}
 C_F (2 C_A - 3 C_F - 4 T_F N_f) & -2 T_F N_f (2 C_A + C_F) \\
 2 C_F (2 C_A + C_F)  & 8\, C_A^{2}-4\,{\it T_F N_f }\,{\it C_F}
 \end{array} \right).
\label{eqP1}
\end{equation}
%-----------------------------------------------------------------------
They agree with the respective leading $\ln^{2l} x $ terms of the 
complete LO and NLO splitting function matrices~\cite{LO,NLO}.
We also list the entries of $\PV^{(2)}_{x\rightarrow 0}$ and
$\PV^{(3)}_{x\rightarrow 0}$ explicitely in the colour factors:
%-----------------------------------------------------------------------
\begin{eqnarray}
P_{qq}^{(2)} &=& \frac{2}{3} C_F \left[
 -5\, C_F^{2}-\frac{3}{2}\, C_A^{2}+6\,{\it C_A}\, C_F
 -8\,{\it T_F N_f }\, C_F-6\,{\it T_F N_f}\,{\it C_A} \right]
\nonumber\\
 P_{qg}^{(2)} &=& \frac{2}{3} T_F N_f \left[
 -15\, C_A^{2}+2\, C_F^{2}-6\,{\it C_F}\,{\it C_A}+8\,T_F N_f C_F 
 \right]
\nonumber\\
 P_{gq}^{(2)} &=& \frac{2}{3} C_F  \left[
 15\, C_A^{2}-2\, C_F^{2}+6\,{\it C_F}\,{\it C_A}-8\,T_F N_f C_F \right]
\nonumber\\
 P_{gg}^{(2)} &=& \frac{2}{3} \left[
 28\, C_A^{3}+2\, T_F N_f\, C_A^{2}- 4\, T_F N_f\, C_F^{2} 
 -24\, C_F\, T_F N_f\, C_A  \right],
\label{eqP2}
\end{eqnarray}
%-----------------------------------------------------------------------
and
%-----------------------------------------------------------------------
\begin{eqnarray}
 P_{qq}^{(3)} &=& \frac{2}{45} C_F \Bigl [
 6\, C_A^{3}-20\, C_F C_A^{2}+22\, C_A C_F^{2}- \frac{19}{2}\, C_F^{3}
 -74\, T_F N_f\, C_A^{2}-44\, C_F T_F N_f\, C_A \nonumber\\
 & &~~~~~~~~~~+2\, T_F N_f\, C_F^{2}+40\, (T_F N_f)^2 C_F \Bigr ] 
\nonumber\\
 P_{qg}^{(3)} &=& \frac{2}{45} T_F N_f  \Bigl [
 -54\, C_F\, C_A^{2}-2\,C_F^{2} C_A+40\,T_F N_f C_F^{2}-128\, C_A^{3}
 -8\, T_F N_f C_A^{2} +15\,C_F^{3} \nonumber\\
 & &~~~~~~~~~~~~~+108\,T_F N_f C_F C_A \Bigr ] 
\nonumber\\
 P_{gq}^{(3)} &=& \frac{2}{45} C_F  \Bigl [
 54\, C_F\, C_A^{2}+2\,C_F^{2} C_A-40\,T_F N_f C_F^{2}+128\, C_A^{3}
 +8\, T_F N_f C_A^{2} -15\,C_F^{3} \nonumber\\
 & &~~~~~~~~~~~~~-108\,T_F N_f C_F C_A \Bigr ] 
\nonumber\\
 P_{gg}^{(3)} &=& \frac{2}{45} \Bigl [
 -288\, T_F N_f\, C_F\, C_A^{2} -64\, C_F^{2} T_F N_f\, C_A
 +6\, T_F N_f\, C_F^{3} +40\, (T_F N_f)^{2} C_F^2 \nonumber\\
 & &~~~~~~+20\, T_F N_f\, C_A^{3} +252\, C_A^{4} \Bigr ].
\label{eqP3}
\end{eqnarray}
%-----------------------------------------------------------------------
Comparing eqs.~(\ref{eqP0}--\ref{eqP3}) it is interesting to note that 
the leading small-$x$ off--diagonal elements are related by
%-----------------------------------------------------------------------
\begin{equation}
 P_{qg}^{(l)}/(T_F N_f) = - P_{gq}^{(l)}/C_F \: .
\end{equation}
%-----------------------------------------------------------------------
We have verified this property 
analytically up to order $l = 100$.~\footnote{We used the program
system {\sf Maple~V}~\cite{MAP}
 for checks and the derivation of higher order 
coefficients given subsequently.}
The numerical values of the matrix elements of $\PV_{x \rightarrow 0}
^{(l)}$ up to $l =10$ are listed in Table~1 for $SU(N_c=3)$, leaving 
the number of (massless) partonic
flavours as the only free parameter
since the corresponding analytical results become rather lengthy. 
The matrix elements for even higher indices are easily obtainable. 
\begin{table}[ht]
\begin{center}
\small
\begin{tabular}{||r|l|l|l|l||}
\hline \hline
\multicolumn{1}{||c|}{ } &
\multicolumn{4}{  c||}{ } \\[-0.4cm]
\multicolumn{1}{||c|}{   } &
\multicolumn{4}{  c||}{$N_f = 3$   } \\ 
\multicolumn{1}{||c|}{ } &
\multicolumn{4}{  c||}{ } \\[-0.4cm] \hline
 & & & & \\[-0.4cm]
\multicolumn{1}{||c|}{$l$} & 
\multicolumn{1}{c|}{$P_{qq}^{(l)}$} & 
\multicolumn{1}{c|}{$P_{qg}^{(l)}$} & 
\multicolumn{1}{c|}{$P_{gq}^{(l)}$} &
\multicolumn{1}{c||}{$P_{gg}^{(l)}$} \\
 & & & & \\[-0.4cm] \hline \hline
 & & & & \\[-0.4cm] 
      0 &  ~0.2666666667D1 & 
	   -6.D0           &
           ~0.5333333333D1 &
           24.D0           \\
      1 &  -0.1066666667D2 &
           -44.D0          &
           ~0.3911111111D2 &
           128.D0          \\
      2 &  -0.3679012346D2 &
           -0.1394444444D3 &
           ~0.1239506173D3 &
           0.4188888889D3  \\
      3 &  -0.6642085048D2 &
           -0.2288296296D3 &
           ~0.2034041152D3 &
           0.6981037037D3  \\
      4 &  -0.6110486315D2 &
           -0.2154469209D3 &
           ~0.1915083742D3 &
           0.6685486038D3  \\
      5 &  -0.3858083350D2 &
           -0.1347153415D3 &
           ~0.1197469702D3 &
           0.4201415122D3  \\
      6 &  -0.1679581048D2 &
           -0.5955393524D2 &
           ~0.5293683133D2 &
           0.1868497637D3  \\
      7 &  -0.5631967112D1 &
           -0.1979044200D2 &
           ~0.1759150400D2 &
           0.6213345892D2  \\
      8 &  -0.1424435573D1 &
           -0.5081773820D1 &
          ~0.4517132284D1  &
           0.1601700122D2  \\
      9 &  -0.2991318204D0 &
           -0.1049686114D1 &
           ~0.9330543234D0 &
           0.3302769905D1  \\
     10 &  -0.4868787168D-1 &
           -0.1756718578D0 &
           ~0.1561527625D0 &
           0.5557315074D0  \\
\hline \hline
\multicolumn{1}{||c|}{ } &
\multicolumn{4}{  c||}{ } \\[-0.4cm]
\multicolumn{1}{||c|}{   } &
\multicolumn{4}{  c||}{$N_f = 4$   } \\ 
\multicolumn{1}{||c|}{ } &
\multicolumn{4}{  c||}{ } \\[-0.4cm] \hline
 & & & & \\[-0.4cm]
\multicolumn{1}{||c|}{$l$} &
\multicolumn{1}{c|}{$P_{qq}^{(l)}$} &
\multicolumn{1}{c|}{$P_{qg}^{(l)}$} &
\multicolumn{1}{c|}{$P_{gq}^{(l)}$} &
\multicolumn{1}{c||}{$P_{gg}^{(l)}$} \\ 
 & & & & \\[-0.4cm] \hline \hline
 & & & & \\[-0.4cm] 
      0 & ~0.2666666667D1  &
        -8.D0              &
        ~0.5333333333D1    &
        ~24.D0             \\
      1 & -16.D0           &
        -0.5866666667D2    &
        ~0.3911111111D2    &
        ~0.1226666667D3    \\
      2 & -0.4953086420D2  &
        -0.1788148148D3    &
        ~0.1192098765D3    &
       ~0.3905185185D3    \\
      3 & -0.8573278464D2  &
        -0.2859456790D3    &
        ~0.1906304527D3    &
        ~0.6315654321D3   \\
      4 & -0.7633439480D2  &
        -0.2595199393D3    &
        ~0.1730132928D3    &
        ~0.5831456986D3   \\
      5 & -0.4648843309D2  &
        -0.1568583694D3    &
        ~0.1045722463D3    &
        ~0.3540380256D3   \\
      6 & -0.1955536907D2  &
        -0.6687789759D2    &
        ~0.4458526506D2    &
        ~0.1519476618D3   \\
      7 & -0.6326464875D1  &
        -0.2148071223D2    &
        ~0.1432047482D2    &
        ~0.4881870016D2   \\
      8 & -0.1545655016D1  &
        -0.5319199339D1    &
        ~0.3546132893D1    &
        ~0.1214527167D2   \\
      9 & -0.3133281583D0  &
        -0.1062889464D1    &
        ~0.7085929761D0    &
        ~0.2420644158D1   \\
     10 & -0.4922813282D-1 &
        -0.1712782960D0    &
        ~0.1141855307D0    &
        ~0.3928177051D0   \\
\hline \hline
\end{tabular}
\normalsize

\vspace{3mm}
\noindent
{\sf Table~1:~~The elements of the coefficient matrices $\PV_{x 
\rightarrow 0}^{(l)}$ in eq.~(\ref{eq31}) for $N_f = 3$ and $N_f = 4$.}
\end{center}
\vspace{-5mm}
\end{table}

We have also calculated the matrices $\PV^{(l)}_{x \rightarrow 0}$ 
for an ${\cal N} = 1$ supersymmetric Yang--Mills field theory, i.e.\ 
$C_A = C_F = 1$, $N_f = 1, T_F = 1/2$. One finds that in this 
case the so--called supersymmetric relation
%-----------------------------------------------------------------------
\begin{equation}
 P_{qq}^{(l)}(x)+ P_{gq}^{(l)}(x)- P_{qg}^{(l)}(x)- P_{gg}^{(l)}(x) = 0
\end{equation}
%-----------------------------------------------------------------------
is fulfilled for the small-$x$ leading terms. We have verified this 
behaviour explicitely up to order $l = 100$. Beginning with 
$O(\alpha_s^2)$ even
%-----------------------------------------------------------------------
\begin{equation}
 P_{qq}^{(l)} - P_{qg}^{(l)} = 0 \: , \:\:\:
 P_{gq}^{(l)} - P_{gg}^{(l)} = 0
\end{equation}
%-----------------------------------------------------------------------
holds, and the matrix of the small-$x$ transition functions depends only
on one single scalar coefficient $p_l$ at each order in $\alpha_s$. 
Hence one can write
%-----------------------------------------------------------------------
\begin{equation}
\label{eqSU}
 \PV^{{\rm SUSY}}_{x \rightarrow 0} = 2\, a_s \MV_1
 + \sum_{l=1}^{\infty} a_s^{l+1} \ln^{2l} x\, p_l \MV_2
\end{equation}
%-----------------------------------------------------------------------
with
%-----------------------------------------------------------------------
\begin{equation}
 \MV_1 \equiv \MV_0^{{\rm SUSY}} =
 \left ( \begin{array}{rr} 1 & -1 \\ 2 & 4 \end{array} \right ),~~~~~ 
 \MV_2 = \left( \begin{array}{rr} -1 & -1 \\ 2 & 2 \end{array} \right ).
\end{equation}
%-----------------------------------------------------------------------
Note that there is a series of relations between the matrices $\Mvec_1$
and $\MV_2$:
%-----------------------------------------------------------------------
\begin{equation}
\Mvec_1 \Mvec_2 = 3 \Mvec_2,~~\Mvec_2^2 = \Mvec_2,~~\left [
\Mvec_1, \Mvec_2 \right ] = 0,
\end{equation}
%-----------------------------------------------------------------------
which determines the structure of (\ref{eqSU}), and a similar
result is obtained for $\Fvec^{\rm SUSY}_{8}(N)$.

We have furthermore investigated whether the dominant terms found in the
present small--$x$ expansion can be obtained in the large--$N_f$
expansion~\cite{GRACEY} partly, where predictions have been obtained on 
the behaviour of the  eigenvalue
%-----------------------------------------------------------------------
\begin{equation}
 e_2 = \frac{1}{2} \left [ P_{qq}(N) + P_{gg}(N) \right ] -
       \frac{1}{2} \sqrt{\left [ P_{qq}(N) - P_{gg}(N) \right ]^2 +
                              4 P_{qg}(N) P_{gq}(N)} \:
\end{equation}
%-----------------------------------------------------------------------
of the Mellin transformed all--order  singlet
transition matrix.
However, as it is found for the lowest order in $\alpha_s$ already, the
contributions to the large $N_f$ expansion of $e_2$ can be obtained only
by accounting for non--leading terms for $x \rightarrow 0$.
Therefore the limits $x \rightarrow 0$ and $N_f \rightarrow \infty$
do not interchange and a further test on the elements of $\PV^{(l)}(N)$
can not be obtained in this way\footnote{Recall that  the non-singlet
transition functions in the small-$x$ limit do not depend on
$N_f$ at all \cite{BV1}.}.
%
%
%%%%%%%%%%%%%%%%%%%%%%%%%%%%%%%%%%%%%%%%%%%%%%%%%%%%%%%%%%%%%%%%%%%%%%%
\section{Numerical results}
%%%%%%%%%%%%%%%%%%%%%%%%%%%%%%%%%%%%%%%%%%%%%%%%%%%%%%%%%%%%%%%%%%%%%%%
%
%
After transformation to Mellin moments, the singlet evolution equation
(\ref{eq21}) is reduced to a system of coupled ordinary differential
equations. Unlike in the non--singlet case considered in refs.\ 
\cite{BV1,BV2} the solution cannot be given in a closed analytical form 
beyond LO here, due to the non--commutativity of the matrices
$\PV^{(i)}(N)$ for different orders in $a_s$. Instead the evolution 
taking into account the leading small-$x$ resummed kernels (\ref{eqSER})
has to be written down in terms of a power series in $a_s$, yielding 
%-----------------------------------------------------------------------
\begin{equation}
\label{SOL}
  \left ( \begin{array}{c}  \Delta \Sigma(N,a_s) \\ \Delta g(N, a_s)
  \end{array} \right )
 =\left [ 1 + \sum_{i=1} a_s^i \UV^{(i)}(N) \right ]
  \left( \frac{a_s}{a_0} \right)^{-{\small \PV}^{(0)}(N)/\beta_0}
  \left[ 1 + \sum_{i=1} a_0^i \UV^{(i)}(N) \right]^{-1}
  \left ( \begin{array}{c}  \Delta \Sigma(N,a_0) \\ \Delta g(N, a_0)
\end{array} \right )
\end{equation}
%-----------------------------------------------------------------------
with $a_0 = a_s(Q^2_0)$. The singlet evolution matrices $\UV^{(i)}(N)$ 
can be expressed in terms of the splitting function moments $\PV^{(j<i)}
(N)$. Technical details can be found elsewhere~\cite{BRV}.
Due to the structure of eq.~(\ref{eq31}), the solution (\ref{SOL}) is 
necessarily related to an asymptotic expansion. For all practical 
cases, say $ x > 10^{-6} $, however, retaining 8 -- 10 terms in 
eq.~(\ref{SOL}) is adequate for obtaining accurate and stable results. 
The transformation of the outcome back to $x$-space finally affords one
standard numerical integration in the complex $N$-plane.

We study the numerical consequences of the resummation (\ref{eq31}) 
using the recent NLO parametrization of ref.~\cite{GRSV} (GRSV) as the 
input at the reference scale $ Q_{0}^{2} = 4 \mbox{ GeV}^2 $. The 
presently large ambiguities due to the virtually unknown polarized gluon
distribution will be briefly illustrated by the maximal and minimal 
$\Delta g$ scenario of the same group. For a short review on current 
parametrizations of polarized parton densities see, e.g., ref.\ 
\cite{LAD}.
The number of flavours $N_f$ in the $\beta $-function is fixed at
$ N_f = 4 $ for all results shown below, and $ \Lambda \equiv \Lambda_
{\overline {\rm MS}} (N_f = 4) = 200 \MeV$ is employed in the standard
approximation to the NLO running of $a_s$, 
%-----------------------------------------------------------------------
\begin{equation}
 a_s(Q^2) = \frac{1}{\beta_0 \ln(Q^2/\Lambda^2)} \left [ 1 - \frac{
 \beta_1}{\beta_0^2} \frac{\ln \ln(Q^2/\Lambda^2)} {\ln(Q^2/ \Lambda^2)}
 \right ] \: .
\end{equation}
%-----------------------------------------------------------------------
The number of partonic flavours in the splitting functions 
$\PV^{(i)}(N)$ is restricted to $N_f^{\prime}=3 $ \cite{GRSV,GRV94}.

We are now ready to present the resummation effects on the polarized 
singlet quark and gluon densities, $x\Delta \Sigma $ and $x\Delta g $,
as well as on the proton and neutron structure functions, $g_1^{\, p}$ 
and $g_1^{\, n}$, respectively. The structure functions have been
obtained from the  (scheme-dependent) parton distributions by the 
convolution (\ref{eq23}), i.e.\ without subtracting the resulting 
subleading $a_s^2$ terms. The non--singlet case has been studied for 
our present input choice in ref.~\cite{BV2}. The corresponding 
resummation effect is negligible (about 1\% or less) over the full 
kinematical region considered here.
In Figure~1 the NLO and leading small-$x$ resummed ($Lx$) results are
displayed for the singlet parton densities, and Figure~2 depicts the
corresponding proton and neutron structure functions $g_1^{\, p,n}$. 

The resummation effects are much larger than for the non--singlet
quantities, as to be expected from the comparison of the coefficients
in Table~1 to the corresponding non--singlet results. E.g., the ratio
{\em (NLO + Lx)/(NLO)} amounts to about 1.72 (1.64) for $\Delta
\Sigma $ ($\Delta g $), respectively, at $Q^2 = 10 \mbox{ GeV}^2 $ and
$ x = 10^{-4} $. 
It should be noted in this context that the small-$x$ evolution strongly
depends on the practically unknown gluon input distribution. This is 
illustrated in Table~2 where the resummed results of Figure~1 are
compared at two representative values of $x$ and $Q^2$ to those obtained
by evolving in the same way the `minimal $\Delta g $' and `maximal 
$\Delta g$' distributions of ref.~\cite{GRSV}. One finds variations up 
to a factor of almost 5 (10) for $\Delta \Sigma $ ($\Delta g $), 
respectively, indicating that at present the input ambiguities are
the dominant source of uncertainties also at small $x$.
\begin{table}[htb]
\begin{center}
\begin{tabular}{||c|c|c|c|c||}
\hline \hline
\multicolumn{1}{||c}{ } &
\multicolumn{2}{|c|}{ } &
\multicolumn{2}{c||}{ } \\[-0.4cm]
\multicolumn{1}{||c}{$Q^2$                       } &
\multicolumn{2}{|c|}{$      10  \mbox{ GeV}^2$   } &
\multicolumn{2}{c||}{$      100 \mbox{ GeV}^2$   } \\
\multicolumn{1}{||c}{ } &
\multicolumn{2}{|c|}{ } &
\multicolumn{2}{c||}{ } \\[-0.4cm]
\hline
 & & & & \\[-0.4cm]
\multicolumn{1}{||c|}{$x$} &
\multicolumn{1}{c|} {$10^{-4}$} &
\multicolumn{1}{c|} {$10^{-3}$} &
\multicolumn{1}{c|} {$10^{-4}$} &
\multicolumn{1}{c||}{$10^{-3}$} \\
 & & & & \\[-0.4cm]
\hline \hline
 & & & & \\[-0.4cm]
                 & -0.0100  & -0.0169  & -0.0171  & -0.0218  \\
$x \Delta\Sigma $& -0.0285  & -0.0396  & -0.0505  & -0.0523  \\
                 & -0.0473  & -0.0560  & -0.0855  & -0.0772  \\
\hline
 & & & & \\[-0.4cm]
                 &  0.019   &  0.034   &  0.053   &  0.071   \\
$x \Delta g $    &  0.101   &  0.152   &  0.226   &  0.281   \\
                 &  0.201   &  0.294   &  0.432   &  0.528   \\
\hline \hline
\end{tabular}

\vspace{2mm}
\end{center}
{\sf Table~2:~A comparison of the resummed evolution of the polarized 
parton distributions for different assumptions on the gluon distribution
$\Delta g$. Upper lines: minimal gluon, middle lines: standard set, lower 
lines: maximal gluon (and corresponding quark distributions) of 
ref.~\protect\cite{GRSV} at $ Q_0^2 = 4 \mbox{ GeV}^2 $.}
\end{table}

An obvious question concerning the large effects found is whether the
resummed $a_s^{l+1} \ln ^{2l} x$ terms really dominate with respect to
the presently yet uncalculated terms less singular in $\ln x$. Recall 
that we found for the non--singlet structure functions that this is 
not the case \cite{BV1,BV2}.
Lacking any further information on the higher--order splitting 
functions, e.g. from sum rules as in the non--singlet and unpolarized 
singlet cases, it appears reasonable to assume that the coefficients of
the first less singular term is of roughly the same size but of
opposite sign as the leading one. To obtain a first estimate we use
\begin{equation}
  \Delta P^{(i>1)} \rightarrow \Delta P^{(i>1)} \cdot (1-N) \: .
\end{equation}
This assumption is motivated by corresponding relations in the
LO and NLO splitting functions where, e.g., for the largest quantity 
$P_{gg}$:
\begin{equation}
  P^{(0)}_{gg}(N) = \frac{24}{N} - 15 + {O}(N)  \: , \:\:\:
  P^{(1)}_{gg}(N) = \frac{256}{N^3} - \frac{244}{N^2} + {O}(1/N)
\end{equation}
for three flavours.
The corresponding results are also given in Figures~1 and~2, and are
denoted by $Lx * (1-N)$. One finds that the corrections due to these
new contributions are very sizeable. In the $x$-range considered here
the effect of the $Lx$--resummation is practically cancelled. Hence the
calculation of also the less singular terms in the higher--order 
splitting functions is indispensable for arriving at sound conclusions 
on the small $x$ evolution on the small-$x$ polarized evolution.
%
%
%%%%%%%%%%%%%%%%%%%%%%%%%%%%%%%%%%%%%%%%%%%%%%%%%%%%%%%%%%%%%%%%%%%%%%%
\section{Conclusions}
%%%%%%%%%%%%%%%%%%%%%%%%%%%%%%%%%%%%%%%%%%%%%%%%%%%%%%%%%%%%%%%%%%%%%%%%
%
%
We have investigated the effect of the resummation of terms of order 
$\alpha_s^{l+1} \ln^{2l} x$, derived from the infrared evolution 
equations in \cite{BER2}, on the small-$x$ behaviour of polarized 
singlet parton distributions and the structure function $g_1$ in 
deep--inelastic scattering. The comparison with the corresponding 
contributions obtained in the same order by complete NLO calculations 
of the splitting function matrix \cite{NLO} shows the equivalence of 
both approaches in this limit up order $\alpha_s^2$. Since the 
coefficient functions up to two--loop order contain only terms less 
singular in $\ln x$ in the $\overline {\mbox{MS}} $ scheme, the 
contributions $\propto \alpha_s^3 \ln^4 x$ 
in the three--loop $\overline
{\mbox{MS}}$ splitting functions $\PV^{(2)}(x)$ have been predicted on 
the basis of this resummation.

As a general result for $SU(N_c)$, the off--diagonal elements of the
matrix $\PV(x,\alpha_s)_{x \rightarrow 0}$ 
are found to be proportional by
the factor $-T_F N_f/C_F$.
The supersymmetric relation holds for the leading small-$x$ terms of 
the anomalous dimensions in ${\cal N} = 1$ supersymmetric Yang--Mills
theory. Starting with order $l=1$ even more constraining relations are 
found in this case and all anomalous dimensions are related at a given 
order in $\alpha_s$.

The numerical analysis shows that the all--order resummation of the
terms $O(\alpha^{l+1}_s \ln^{2l} x)$ leads to very large corrections at 
small $x$. Terms less singular in $\ln x $, being not calculated yet,
can be expected to contribute in a very significant way even at the
smallest $x$-values considered, $ x\simeq 10^{-5} $, as in the case of
leading and next--to--leading order. Even a full
compensation of the effect
obtained resumming the most singular terms can not be excluded.
Hence solid conclusions on the 
small-$x$ evolution of polarized singlet parton densities and structure 
functions can only be drawn if these terms are calculated.

\vspace{5mm}
\noindent
{\bf Acknowledgements}
We would like to thank the authors of ref.~\cite{GRSV} for providing 
the input parameters of their maximal and minimal $\Delta g$ sets.
This work was supported in part by the German Federal Ministry for 
Research and Technology under contract No.\ 05 7WZ91P (0).
%
%%%%%%%%%%%%%%%%%%%%%%%%%%%%%%%%%%%%%%%%%%%%%%%%%%%%%%%%%%%%%%%%%%%%%%%%

\vspace{1cm}

%%%%%%%%%%%%%%%%%%%%%%%%%%%%%%%%%%%%%%%%%%%%%%%%%%%%%%%%%%%%%%%%%%%%%%%%
\noindent
{\Large\bf Figures}
%%%%%%%%%%%%%%%%%%%%%%%%%%%%%%%%%%%%%%%%%%%%%%%%%%%%%%%%%%%%%%%%%%%%%%%%

\vspace{0.5cm}
\noindent
{\sf Figure~1: The $Q^2$ evolution of the polarized singlet quark and 
gluon momentum distributions $x\Delta \Sigma$ and $x\Delta g$ as 
obtained from the GRSV standard distribution \cite{GRSV} at $Q_0^2 = 4 
\GeV^2$. The results are shown for the NLO kernels (full), the leading
small-$x$ resummed kernels (dashed), and a modification of the latter
by possible less singular terms discussed in the text (dotted curves).}

\vspace{0.5cm}
\noindent
{\sf  Figure~2: The $x$ and $Q^2$ behaviour of the polarized proton and
neutron structure functions $g_1^{\, p,n}(x,Q^2)$ as obtained from the 
parton densities in the previous figure. The notations are the same as 
in Figure~1.}

\newpage
\vspace*{-1.0cm}
\hspace*{0.6cm}
\epsfig{file=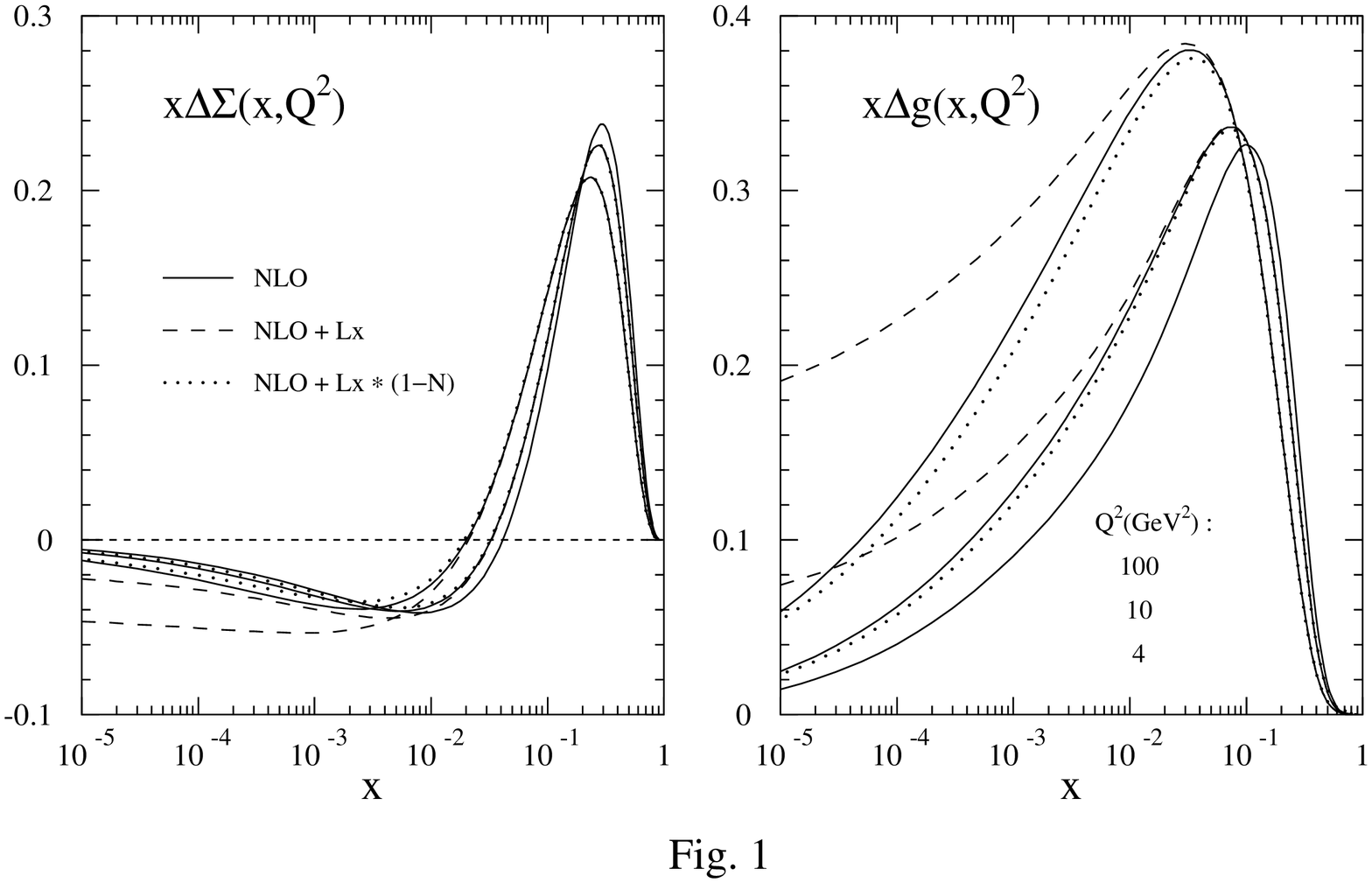,height=15.5cm,angle=90}

\newpage
\vspace*{-1.0cm}
\hspace*{0.6cm}
\epsfig{file=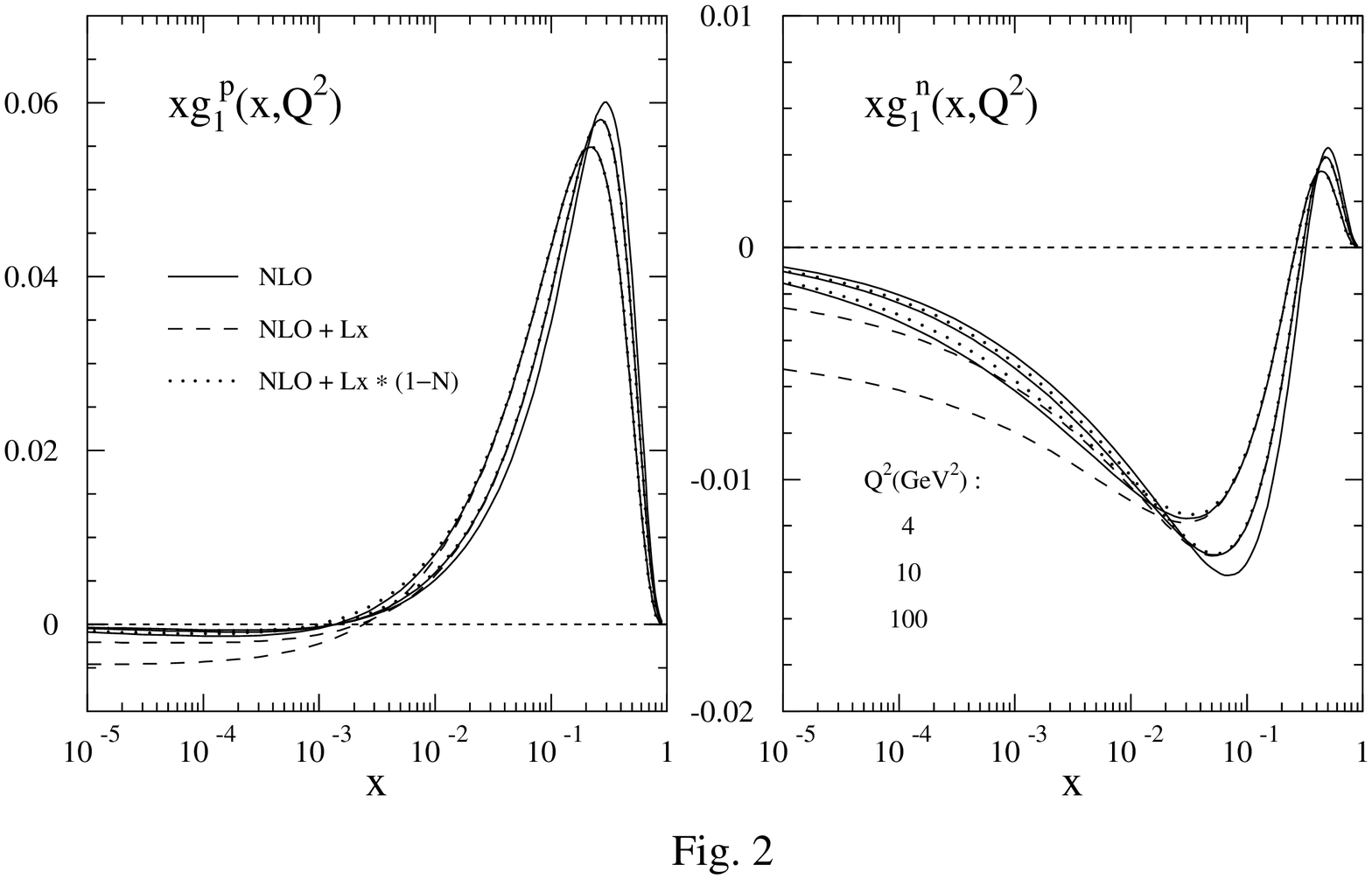,height=15.5cm,angle=90}

% av final to here

\end{document}